\documentclass[12pt]{article}
\usepackage[cp1251]{inputenc}
\usepackage{amsfonts,amsmath}
\newtheorem{theorem}{Theorem}

\begin{document}

\title {\hbox{\normalsize }\hbox{}
Two-dimensional discrete operators and rational functions on algebraic curves
}

\author {Polina~A.~Leonchik, Andrey~E.~Mironov}

\date{}
\maketitle

\sloppy

\begin{abstract}
In this paper we study a connection between finite-gap on one energy level two-dimensional Schr\"odinger operators and two-dimensional discrete operators. We find spectral data for a new class of two-dimensional integrable discrete operators. These operators have eigenfunctions on zero level energy parameterized by points of algebraic spectral curves. In the case of genus one spectral curves we show that the finite-gap Schr\"odinger operators can be obtained as a limit of the discrete operators.
\end{abstract}

\section{Introduction and main results}

In \cite{DKN} Dubrovin, Krichever and Novikov introduced a remarkable class of periodic  finite-gap on one energy level Schr\"odinger operators
\begin{gather*}
    H = \left(i \partial_x - A_1 \right)^2 + \left(i \partial_y - A_2 \right)^2 + u(x,y), \quad A_j = A_j (x,y), 
\end{gather*}
where the potential $u$ and the vector-potential $(A_1, A_2)$ are periodic in $x$ and $y$ with periods $T_1$ and $T_2$. Bloch functions $\psi(x,y)$ of $H$, i.e. such functions that
\begin{gather*}
    H \psi = 0,\quad
    \psi (x+T_1, y) = e^{i p_1 T_1} \psi (x,y) , \quad \psi (x, y+T_2) = e^{i p_2 T_2} \psi (x,y), 
\end{gather*}
are parameterized by a Riemann surface $\Gamma$ of genus $g$, $\psi=\psi(x,y,P), P\in\Gamma.$ Operator $H$ and function $\psi$ are reconstructed from spectral data that contains the Riemann surface $\Gamma$, two marked points $p,q \in \Gamma$, where $\psi$ has essential singularities and a non-special divisor of degree $g$ on $\Gamma$ (see section 2). Coefficients of $H$ and $\psi$ are expressed in terms of the theta-function of the Jacobi variety of $\Gamma$. Function $\psi$ is called {\it the two-point Baker--Akhiezer function} (see \cite{DKN}). Many problems in geometry and mathematical physics were later  solved with the help of the Baker--Akhiezer function and its generalizations.
 For example, constant mean curvature tori in ${\mathbb R}^3$ \cite{Bob}, minimal Lagrangian tori in ${\mathbb C}P^2$ \cite{Sharip}, \cite{CM}, orthogonal curvilinear coordinate systems in ${\mathbb R}^n$ \cite{K2} are described in terms of the Baker--Akhiezer functions. Novikov and Veselov \cite{NV} found spectral data corresponding to potential Schr\"odinger operators, i.e. $A_1=A_2=0$. In \cite{GMN1}, \cite{GMN2} completely factored reduction $H=-(\partial+A)(\bar{\partial}-\bar{A})$ was found.
 Feldman, Knorrer, Trubowitz \cite{FKT} proved that the periodic Schr\"odinger operator can be finite-gap only on one energy level, excluding case with potentials of the type $u(x,y)=u_1(x)+u_2(y)$, where both $u_1(x)$ and $u_2(y)$ are finite-gap. In other words Bloch functions on the energy level $E \neq 0, H \psi = E \psi$ form Riemann surface of infinite genus. 

In \cite{K} Krichever constructed a discrete analogue of the theory of the finite-gap Schr\"odinger operators. The direct and the inverse problems for discrete operators of the form
\begin{gather}
    L \psi (n,m) = \psi(n+1, m+1) + a_{n,m} \psi (n+1, m) + \notag \\
    + b_{n,m} \psi (n,m+1) + v_{n,m} \psi(n,m)  \label{eq1}
\end{gather}
with periodic coefficients 
$$ 
 a_{n,m} = a_{n+N, m} = a_{n,m+M}, \quad b_{n,m} = b_{n+N, m} = b_{n,m+M}, 
 $$
 $$ v_{n,m} = v_{n+N, m} = v_{n,m+M}, n, m \in \mathbb Z, $$ 
were solved. Here $N,M\in\mathbb{N}$ are periods.

The equation $L \psi = 0$ can be obtained as a 4-point scheme. A. Doliwa, P. Grinevich, M. Nieszporski, P.M. Satini in \cite{DGNS} considered discrete Moutard reductions and algebro-geometric solutions of a specific 4-point scheme with some constraints on the spectral data. The constraints which give rise to the studied 4-point scheme are analogous to those in the study of the two-dimensional continuous Schr\"odinger operators. 

The main results of this paper are the following. We find new spectral data for operators of the form \eqref{eq1} (non-periodic) such that eigenfunctions $\psi$, $L\psi=0$ are parameterized by algebraic spectral curves. We find a connection between the Schr\"odinger operator $H$ and discrete operator $L$. In the case of spectral curves of genus one we show that $H$ can be obtained as some limit of $L$. Similar results in the one-dimensional case were obtained in \cite{MM}. More precisely, in \cite{MM} it is shown that commuting ordinary rank one differential operators can be extended to commuting difference operators preserving all possible integrable properties, such as the spectral curve etc.

In section 2 we remind how to reconstruct finite-gap on one energy level two-dimensional Schr\"odinger operators from spectral data \cite{DKN}. In section 3 we find spectral data for a new class of integrable discrete operators $L$. The main result of this paper is Theorem 1. We prove that using spectral data one can reconstruct a unique function $\psi(n,m,P), P\in\Gamma$ and an operator $L$, such that $L\psi(n,m,P)=0$. For fixed $n$ and $m$ function $\psi$ is a rational function on $\Gamma$ with some special zero and pole divisor. In section 4 in the case of elliptic spectral curve we extend $L$ to a difference operator $L_{\varepsilon, \delta}$ depending on small parameters $\varepsilon, \delta$. We find  conditions  for $\lim\limits_{\varepsilon, \delta \to 0} L_{\varepsilon, \delta} = H$.

\section{Finite-gap Schr\"odinger operators}

In this section we remind the construction of finite-gap on one energy level two-dimensional Schr\"odinger operators \cite{DKN}. Let us take the following spectral data
$$
 S=\{\Gamma,\gamma,p,q\},
$$
where $\gamma=\gamma_1+\ldots+\gamma_g$ is a non-special divisor of degree $g$ on the Riemann surface $\Gamma$ of genus $g$,
$p,q\in\Gamma$ are marked points. There is a unique function $\psi(z,\bar{z},P), P\in\Gamma$ having the following properties:

1. In the neighbourhood of $p$ and $q$ $\psi$ has the form

$$
 \psi=e^{zk_1}\left(1+\frac{\xi(z,\bar{z})}{k_1}+\ldots\right),
$$
$$
\psi=e^{\bar{z}k_2}\left(c(z,\bar{z})+\frac{\eta(z,\bar{z})}{k_2}+\ldots\right),
$$
where $k_1^{-1}, k_2^{-1}$ are local parameters on $\Gamma$ near $p$ and $q$.

2. On $\Gamma \backslash \{p, q\}$ the function $\psi$ is meromorphic with the pole divisor $\gamma$.

The function $\psi$ is called {\it the two-point Baker--Akhiezer function} \cite{DKN}. From the existence and uniqueness of the function it follows that $\psi$ satisfies the Schr\"odinger equation
$$
 H\psi= (\partial_z \partial_{\bar{z}} + A(z,\bar{z}) \partial_{\bar{z}} + u(z,\bar{z})) \psi=0,
$$
where 
$$
 A = -\partial_z\log c(z,\bar{z}), \quad u(z,\bar{z}) = - \partial_{\bar{z}} \xi(z,\bar{z}).
$$
The potential $u(z,\bar{z})$ has the form
$$
u(z,\bar{z}) = 2 \partial_z\partial_{\bar{z}} \log \theta (U_1 z + U_2 \bar{z} + U_3),
$$
where $\theta$ is the theta-function of the Jacobi variety $J(\Gamma)$ of $\Gamma$, 
$U_1, U_2, U_3$ are some vectors.

In the case of an elliptic spectral curve the operator $H$ and the Baker-Akhiezer function $\psi$ can be found in terms of Weierstrass elliptic functions. Let $\Gamma = \mathbb C / \Lambda$, $\Lambda = \{ 2 m \omega + 2 n \omega', n,m \in \mathbb Z \}$ be the elliptic curve. Weierstrass $\wp$-function is a meromorphic function on $\Gamma$ with a unique pole of  order 2 at $0 \in \Gamma$. It can be defined by the following series
\begin{gather*}
    \wp (w) = \cfrac{1}{w^2} + \sum\limits_{(n,m) \in \mathbb Z ^2 \backslash \{ 0, 0\} } \left( \cfrac{1}{(w - 2 m \omega - 2 n \omega')^2} - \cfrac{1}{(2 m \omega + 2 n \omega' )^2} \right).
\end{gather*}
The $\zeta$- and $\sigma$-function are determined by the following way
\begin{gather*}
    \zeta'(w) = - \wp (w), \qquad
    \cfrac{\sigma'(w)}{\sigma(w)} = \zeta(w).
\end{gather*}
The $\zeta$-function has simple poles at $w = 2 m \omega + 2 n \omega', n,m \in \mathbb Z$ and has the following property 
\begin{gather*}
    \zeta (w + 2 n \omega + 2 m \omega') = \zeta (w) + 2 n \eta + 2 m \eta', 
\end{gather*}
where $\eta = \zeta (\omega), \eta' = \zeta (\omega')$.

The $\sigma$-function is an entire function on $\mathbb C$, it has simple zeros at $w = 2 m \omega + 2 n \omega', n,m \in \mathbb Z$ and  has the following properties
\begin{gather*}
    \sigma(w + 2 \omega) = - \sigma (w) \exp{ (2 \eta (w + \omega)) }, \ \sigma(w + 2 \omega') = - \sigma(w) \exp{ (2 \eta' (w + \omega')) }.
\end{gather*}

Let us take the following spectral data 
$$
 S=\{ \Gamma, \gamma, p, q\}.
$$

Then the Baker--Akhiezer function has the form
\begin{gather*}
    \psi(z, \bar{z}, w) = e^{z \zeta (w - p) + \bar{z} \zeta (w - q)} \cfrac{\sigma (w - z - \bar{z} - \gamma) \sigma (p - \gamma) }{ \sigma (w - \gamma) \sigma (p - z - \bar{z} - \gamma)} e^{- \bar{z} \zeta (p - q)}.
\end{gather*}

The function $\psi$ satisfies the Schr\"odinger equation $H \psi = 0$, where
\begin{gather}
    H = \partial_z\partial_{\bar{z}} + (\zeta (p - q) + \zeta (q - z - \bar{z} - \gamma) - \zeta (p - z - \bar{z} - \gamma) )  \partial_{\bar{z}} + \notag \\ + \wp (p - q) - \wp (p - z - \bar{z} - \gamma). \label{H}
\end{gather}

\section{Discrete analogue of Schr\"odinger operator}

In this section we construct a two-dimensional discrete analogue of the Schr\"odinger operator. The eigenfunctions of this operator are parameterised by points of a Riemann surface.

Let us take the following spectral data
\begin{gather} 
     \widetilde{S} = \{  \Gamma, p, q, \gamma, P_n, Q_m \}, n,m \in \mathbb Z \backslash \{ 0 \}, 
\end{gather}
where $\Gamma$ is a compact Riemann surface of genus $g$, $p, q$ are two marked points on $\Gamma$,  $\gamma = \gamma_1 + \ldots + \gamma_g$ is a non-special divisor of degree $g$ on $\Gamma$,  $P_n, Q_m$ are two infinite families of points on $\Gamma$ in general position.

We introduce the following divisors
\begin{gather*}
    P(n) = 
    \begin{cases}
        P_1 + \ldots + P_n,\  n > 0, \\
        - P_{-1} - \ldots - P_{n},\  n < 0, \\
        0,\  n = 0.
    \end{cases}
    Q (m) = 
    \begin{cases}
        Q_1 - \ldots +Q_m,\ m > 0, \\
        -Q_{-1} - \ldots -Q_{m},\  m < 0, \\
        0,\ m = 0.
    \end{cases}
\end{gather*}

We have the following theorem.

\begin{theorem}
    There exists a unique meromorphic function $\psi(n,m, P)$ on $\Gamma$, $n,m \in \mathbb Z, P \in \Gamma$ with the following properties.

\begin{enumerate}
    \item[(1)] The zero and pole divisor of $\psi$ has the form
    $$ \left( \psi(n,m, P) \right) = P(n) + Q(m) + \gamma (n,m) - n p - m q - \gamma ,$$
where $ \gamma (n,m) = \gamma_1(n,m) + \ldots + \gamma_g(n,m)$ is some divisor on $\Gamma$, $\gamma (0,0) = \gamma$. 

    \item[(2)] In the neighbourhood of $p$ function $\psi(n,m,P)$ has the form 
$$\psi(n,m, P) = k_1^n + O(k_1^{n-1}), $$
where $k_1^{-1}$ is a local parameter in the neighbourhood of $p$.

    \item[(3)] $\psi (0,0,P) = 1$.
\end{enumerate}

Moreover, the function $\psi(n,m,P)$ satisfies the following equation 
$$   L \psi (n,m,P) = \psi(n+1, m+1,P) + a_{n,m} \psi (n+1, m,P) + $$
$$ + b_{n,m} \psi (n,m+1,P) + v_{n,m} \psi(n,m,P) = 0 , $$

where $a_{n,m}, b_{n,m}, v_{n,m}$ are some coefficients.
\end{theorem}

Thus, the operator $L$ has an infinite family of functions from the kernel that is parameterized by points of $\Gamma$. 

{\bf Proof.} Let us consider the case $n > 0, m > 0$ (other cases are similar). By the Riemann-Roch theorem, the dimension of the space of meromorphic functions on $\Gamma$ with pole divisor $ D = n p + m q + \gamma_1 + \ldots + \gamma_g$  is equal to
$$ l(D) = n + m + 1.$$
The requirement that a function from this space vanishes at points
$P_1, \ldots, P_n$, $Q_1, \ldots, Q_m$ extracts the one-dimensional subspace of this space. Condition (2) gives us the unique function $\psi$. 

Let us consider the function 
$$
 \widetilde{\psi}(n,m) = \psi(n+1, m+1) + a_{n,m} \psi (n+1, m) + b_{n,m} \psi (n,m+1).
$$ 
We assume that $n>0, m>0$ (other cases are similar). Choose the coefficients  $a_{n,m}, b_{n,m}$ such that the function $\widetilde{\psi}(n,m)$ has poles of order $n$ at $p$ and order $m$ at $q$. Then the zero and pole divisor of $\widetilde{\psi} (n,m)$ has the form
$$ \left( \widetilde{\psi}(n,m, P) \right) = P(n) + Q(m) + \gamma (n,m) - n p - m q - \gamma.$$
From the uniqueness of $\psi$ it follows that $\widetilde{\psi} (n,m)$ is proportional to $\psi(n,m)$ with some coefficient $- v_{n,m}$
$$
 \widetilde{\psi}(n,m) = - v_{n,m} \psi(n,m).
$$
Hence, we have $L\psi(n,m,P)=0$.
Theorem 1 is proved. 

Let $k^{-1}_2$ be a local parameter in the neighbourhood of $q$. Then we have 
$$
  \psi(n,m, P) = \lambda_{n,m}k_2^m + O(k_2^{m-1}), 
$$
where $\lambda_{0,0} = 1$. Let us introduce the following functions
$$
  \chi_1(n,m, P)=\frac{\psi(n+1,m,P)}{\psi(n,m,P)}, \quad \chi_2(n,m, P)=\frac{\psi(n,m+1,P)}{\psi(n,m,P)}.
$$
These functions have the following properties.

1. The zero and pole divisors of $\chi_1(n,m,P)$ and $\chi_2(n,m,P)$ have the form
\begin{gather}
    \left( \chi_1(n,m, P) \right) = P_{n+1} + \gamma(n+1,m) - p - \gamma(n,m), \quad n \neq -1, \label{div_chi_1} \\
    \left( \chi_1(-1,m, P) \right) = P_{-1} + \gamma(0,m) - p - \gamma(-1,m), \notag \\
    \left( \chi_2(n,m, P) \right) = Q_{m+1} + \gamma(n,m+1) - q - \gamma(n,m), \quad m \neq -1, \label{div_chi_2} \\
    \left( \chi_2(n,-1, P) \right) = Q_{-1} + \gamma(n,0) - q - \gamma(n,-1). \notag
\end{gather}

2. In the neighbourhood of $p$ the function $\chi_1(n,m, P)$ is expanded as follows 
\begin{gather*}
    \chi_1(n,m, P) = k_1 + O(1).
\end{gather*}

In the neighbourhood of $q$ the function $\chi_2(n,m, P)$ is expanded as follows 
\begin{gather*}
    \chi_2(n,m, P) = \cfrac{\lambda_{n,m+1}}{\lambda_{n,m}}k_2 + O(1).
\end{gather*}

The compatibility condition for $\chi_1$ and $\chi_2$ has the form
\begin{gather}
    \chi_1(n,m+1) \chi_2(n,m) = \chi_2(n+1,m) \chi_1(n,m). \label{compatibility}
\end{gather}
    
From the equation $L\psi=0$ we get 
\begin{gather}
    \chi_1(n,m+1) \chi_2(n,m) + a_{n,m} \chi_1(n,m) + b_{n,m} \chi_2(n,m) + v_{n,m} = \label{L_chi_1} \\
    = \chi_2(n+1,m) \chi_1(n,m) + a_{n,m} \chi_1(n,m) + b_{n,m} \chi_2(n,m) + v_{n,m} = 0. \label{L_chi_2}
\end{gather}     

Let us express coefficients $a_{n,m}, b_{n,m}, v_{n,m}$ of $L$ through coefficients of the Laurent series of $\chi_1$ and $\chi_2$ in the neighbourhoods of $p$ and $q$.

Let 
\begin{gather*}
    \chi_1(n,m,P) = k_1 + d(n,m) + O(k^{-1}_1), \\
    \chi_2(n,m,P) = c_0 (n,m) + \frac{c_1 (n,m)}{k_1} + O(k_1^{-2})
\end{gather*}
in the neighbourhood of $p$ and 
\begin{gather*}
    \chi_1 (n,m,P) = s_0 (n,m) + \frac{s_1 (n,m)}{k_2} + O(k_2^{-2})
\end{gather*}
in the neighbourhood of $q$.
From \eqref{compatibility} it follows 
\begin{gather*}
    c_0 (n,m) = c_0 (n+1,m),
\end{gather*}
so $c_0 = c_0 (m)$.

Let us consider \eqref{L_chi_2} in the neighbourhood of $p$. We get
\begin{gather*}
    a_{n,m} + c_0 (m) = 0.
\end{gather*}

Let us consider \eqref{L_chi_1} in the neighbourhood of $q$. We get 
\begin{gather*}
    \cfrac{\lambda_{n,m+1}}{\lambda_{n,m}} s_0 (n,m+1) + b_{n,m} \cfrac{\lambda_{n,m+1}}{\lambda_{n,m}} = 0.
\end{gather*}

Hence 
\begin{gather*}
    a_{n,m} = - c_0 (m), \quad   b_{n,m} = - s_0 (n,m+1).
\end{gather*}

Let us consider \eqref{L_chi_2} in the neighbourhood of $p$. It gives us 
\begin{gather*}
    c_1 (n+1, m) + c_0(m) d (n,m) + a_{n,m} d(n,m) + b_{n,m} c_0 (n,m) + v_{n,m} = 0,
\end{gather*}
hence 
\begin{gather*}
    v_{n,m} = - c_1 (n+1,m) - b_{n,m} c_0(m) = - c_1 (n+1,m) + s_0 (n,m+1) c_0 (m).
\end{gather*}

\textbf{Example.} Let us find the operator $L$ in the case of the elliptic spectral curve. In this case the coefficients of $L$, the functions $\chi_1(n,m,P)$ and $\chi_2(n,m,P)$ can be found in terms of the Weierstrass elliptic functions. 

Let us take the following spectral data: 
\begin{gather} 
     \widetilde{S} = \{  \Gamma, p, q, \gamma, P_n, Q_m \}, n, m \in \mathbb Z \backslash \{ 0 \},      \label{SD}
\end{gather}
where $\Gamma = \mathbb C / \Lambda$ is the elliptic spectral curve, $p, q$ are two marked points on $\Gamma$, $\gamma \in \Gamma$, $P_n, Q_m \in \Gamma$ are two infinite families of points in general position.

Let $\gamma(n,m)$ be a divisor from Theorem 1 (see also \eqref{div_chi_1}, \eqref{div_chi_2}). For our convenience we find $P_n$ and $Q_m$ in terms of $\gamma(n,m)$. 
 
Functions $\chi_1(n,m,P)$ and $\chi_2(n,m,P)$ have the following forms 
\begin{gather}
    \chi_1(n,m, w) = \zeta (w - p) - \zeta (w - \gamma(n,m)) +  \notag \\
    + \zeta (\gamma(n+1,m) - \gamma(n,m)) - \zeta (\gamma(n+1,m) - p),  \label{chi_1_zeta} \\
    \chi_2(n,m, w) = \cfrac{\lambda_{n,m+1}}{\lambda_{n,m}} \left( \zeta (w - q) - \zeta (w - \gamma(n,m)) +  \notag\right. \\
    + \left.\zeta (\gamma(n,m+1) - \gamma(n,m)) - \zeta (\gamma(n,m+1) - q) \right).  \label{chi_2_zeta}
\end{gather}

By direct calculations one can check that 
\begin{gather*}
    \chi_1(n,m, \gamma(n+1,m)) = \chi_2 (n,m, \gamma (n,m+1)) = 0.
\end{gather*}

By direct calculations one can also check (see \eqref{div_chi_1}, \eqref{div_chi_2}) that
\begin{gather*}
    \chi_1 (n,m,P_{n+1}) = \chi_2 (n,m, Q_{m+1}) = 0,
\end{gather*}
where
\begin{gather*}
     P_{n+1} = p - \gamma(n+1,m) + \gamma(n,m), \quad
     Q_{m+1} = q - \gamma(n,m+1) + \gamma(n,m). 
\end{gather*}

From the previous identities it follows that $\gamma(n,m)$ must have the form
\begin{gather*}
     \gamma(n,m) = \gamma_1 (n) + \gamma_2 (m),
\end{gather*}
where $\gamma_1(n), \gamma_2 (m)$ are some functions.

From the definition of $\chi_1$ and $\chi_2$, in the neighbourhood of $p$ we have 
\begin{gather*}
    \chi_1(n,m+1, P) \chi_2 (n,m,P) = \cfrac{\psi(n+1, m+1, P)}{\psi(n,m,P)} = k_1 + O(1).
\end{gather*}

Hence, from \eqref{chi_1_zeta}, \eqref{chi_2_zeta} we get 
\begin{gather*}
    \cfrac{\lambda_{n,m+1}}{\lambda_{n,m}} (\zeta (p - q) -  \zeta ( p - \gamma_1 (n) - \gamma_2(m)) + \\ + \zeta (\gamma_2 (m+1) - \gamma_2 (m)) - \zeta (\gamma_1 (n) + \gamma_2(m+1) - q) ) = 1.
\end{gather*}

This allows us to find explicit formulas for coefficients of $L$
\begin{gather*}
    a_{n,m} = -1,  \\
    b_{n,m} = \zeta (p - q) + \zeta (q - \gamma_1(n) - \gamma_2(m+1)) - \\ 
    - \zeta (\gamma_1(n+1) - \gamma_1(n)) - \zeta (p - \gamma_1(n+1) - \gamma_2(m+1)),  \\
    v_{n,m} = \cfrac{\lambda_{n+1,m+1}}{\lambda_{n+1,m}} \left( \wp (p - q) - \wp (p - \gamma_1(n+1) - \gamma_2(m))  \right) - b_{n,m}.
\end{gather*}

\section{Difference analogue of Sch\"odinger operator}

In this section we discuss how to extend the discrete operator $L$ constructed in the previous section to a difference operator of the form
\begin{gather*}
    L_{\varepsilon, \delta} = \cfrac{T_1}{\varepsilon} \cfrac{T_2}{\delta} + a(x,y) \cfrac{T_1}{\varepsilon} + b(x,y) \cfrac{T_2}{\delta} + v(x,y), 
\end{gather*}
where $T_1, T_2$ are the shift operators 
$$
T_1 f(x,y) = f(x + \varepsilon, y), \quad  T_2 f(x,y) = f(x,y + \delta).
$$

The operator $L_{\varepsilon, \delta}$ must have the same properties as the operator $L$. In addition, there must be a limit of $L_{\varepsilon, \delta}$ at $\varepsilon, \delta \to 0$.

The main difficulty to construct $L_{\varepsilon, \delta}$ is the following. One can not extend discrete variables $n,m$ in Theorem 1 to continuous variables $x,y$, because the number of zeros and poles of $\psi (n,m,P)$ on $\Gamma$ depends on $n,m$. On the other hand the number of zeros and poles of $\chi_1$ and $\chi_2$ does not depend on $n$ and $m$. This allows us to extend $\chi_1$ and $\chi_2$ to continuous variables $x,y \in \mathbb C$ with necessary properties in the case of elliptic spectral curves.

Let us take the following spectral data 
\begin{gather*} 
     \widetilde{S} = \{  \Gamma, p, q,  \gamma, P_{x}, Q_{y} \}, \quad x, y \in \mathbb C, 
\end{gather*}
where $\Gamma$ is the elliptic spectral curve, $P_x = P_x (\varepsilon), Q_y = Q_y (\delta) $ are two smooth families of points, such that $P_x (0) = p$, $Q_y (0) = q$. 

So we replace $P_n, Q_m$ in \eqref{SD} with $P_x, Q_y$. The conditions \eqref{div_chi_1}, \eqref{div_chi_2} we replace with 
\begin{gather*}
    \left( \chi_1(x,y, P) \right) = P_{x+ \varepsilon} + \gamma(x+ \varepsilon,y) - p - \gamma(x,y), \\
    \left( \chi_2(x,y, P) \right) = Q_{y + \delta} + \gamma(x,y+ \delta) - q - \gamma(x,y), 
\end{gather*}

We require that the function $\psi(x,y,P)$ satisfies the equations
\begin{gather} \label{psi}
    \cfrac{T_1}{\varepsilon} \psi (x,y, P) - \chi_1 \psi (x,y, P) = 0, \quad \cfrac{T_2}{\delta} \psi (x,y, P) - \chi_2 \psi (x,y, P) = 0.
\end{gather}

The compatibility condition 
\begin{gather*}
    \cfrac{T_1}{\varepsilon} \cfrac{T_2}{\delta} \psi (x,y, P) = \cfrac{T_2}{\delta} (\chi_1 (x,y, P) \psi(x,y, P) ) = \cfrac{T_1}{\varepsilon} (\chi_2 (x,y, P) \psi(x,y, P))
\end{gather*}
gives us
\begin{gather*}
    \chi_1(x,y + \delta, P) \chi_2(x,y, P) = \chi_2(x + \varepsilon,y, P) \chi_1(x,y, P).
\end{gather*}

Functions $\chi_1 (x,y, P)$, $\chi_2 (x,y, P)$ have the form
\begin{gather*}
    \chi_1(x,y, w) = \zeta (w - p) - \zeta (w - \gamma(x,y)) + \\ 
    + \zeta (\gamma(x + \varepsilon,y) - \gamma(x,y)) - \zeta (\gamma(x + \varepsilon,y) - p) ,  \\
    \chi_2(x,y, z) = \lambda (x,y) ( \zeta (w - q) - \zeta (w - \gamma(x,y)) + \\ 
    + \zeta (\gamma(x,y + \delta) - \gamma(x,y)) - \zeta (\gamma(x,y + \delta) - q) ), 
\end{gather*}
where the coefficient $\lambda(x,y)$ can be found from
\begin{gather*}
    \lambda(x,y) ( \zeta (p - q) - \zeta (p - \gamma(x,y)) + \\ 
    + \zeta (\gamma(x,y + \delta) - \gamma(x,y)) - \zeta (\gamma(x,y + \delta) - q) ) = \cfrac{1}{\delta}.
\end{gather*}

Here
\begin{gather*}
    \gamma(x,y, \varepsilon, \delta) = \gamma_1 (x, \varepsilon) + \gamma_2 (y, \delta), \\
     P_{x + \varepsilon} = p - \gamma_1(x+\varepsilon) + \gamma_1(x),  \\
     Q_{y + \delta} = q - \gamma_2(y + \delta) + \gamma_2(y).
\end{gather*}

Due to \eqref{psi} from $L_{\varepsilon, \delta} \psi = 0$ it follows 
\begin{gather*}
    \chi_1(x,y + \delta) \chi_2(x,y) + a(x,y) \chi_1(x,y) + b(x,y) \chi_2(x,y) + v(x,y) = 0, \\
    \chi_2(x + \varepsilon,y) \chi_1(x,y) + a(x,y) \chi_1(x,y) + b(x,y) \chi_2(x,y) + v(x,y) = 0.
\end{gather*}

From here we find the coefficients 
\begin{gather*}
    a(x,y) = -\cfrac{1}{\delta},  \\
    b(x,y) = \zeta (p - q) + \zeta (q - \gamma_1(x) - \gamma_2(y+ \delta)) - \\ - \zeta (\gamma_1(x + \varepsilon) - \gamma_1(x) ) - \zeta (p - \gamma_1(x + \varepsilon) - \gamma_2 (y + \delta)) , \\
    v(x,y) = \lambda(x + \varepsilon, y) ( \wp (p - q) - \wp (p - \gamma_1(x + \varepsilon) -\gamma_2(y) ) ) - \cfrac{b(x,y)}{\delta} . 
\end{gather*}

Let us assume that $\gamma_1(x)$ and $\gamma_2 (y)$ are expanded in a series as follows
\begin{gather*}
    \gamma_1 (x) = \alpha_0(x) + \alpha_1 (x) \varepsilon + \alpha_2 (x) \varepsilon^2 + \ldots,  \\
    \gamma_2(y) = \beta_0(y) + \beta_1 (y) \delta + \beta_2 (y) \delta^2 + \ldots, 
\end{gather*}
where 
\begin{gather*}
    \alpha_0 (x) = x + c \gamma, \quad \alpha_1 (x) = \operatorname{const}, \\
    \beta_0 (y) = y + (1 - c) \gamma, \quad  \beta_1 (y) = \operatorname{const}.
\end{gather*}

We set $x = z, y = \overline{z}$. By direct calculations one can check that 
\begin{gather*}
    \lim\limits_{\varepsilon, \delta \to 0} L_{\varepsilon, \delta} = H,
\end{gather*}
where $H$ is a finite-gap Schr\"odinger operator given by \eqref{H}.

\section*{Funding}

This work was supported by the Russian Science Foundation, project no. 24-11-00281.

\section*{Competing interests}

The authors declare that they have no conflicts of
interest.

P.A. Leonchik, Novosibirsk State University,
Pirogova st. 1, 630090, Novosibirsk, Russia and
Sobolev Institute of Mathematics,
4 Acad. Koptyug avenue, 630090, Novosibirsk, Russia

e-mail: leonchik.2002@mail.ru

A.E. Mironov, Novosibirsk State University,
Pirogova st. 1, 630090, Novosibirsk, Russia and
 Sobolev Institute of Mathematics,
4 Acad. Koptyug avenue, 630090, Novosibirsk, Russia 

e-mail: mironov@math.nsc.ru

\end{document}